\documentclass[journal]{IEEEtran}

\usepackage{graphicx}
\usepackage{placeins}

\ifCLASSINFOpdf
\else
\fi

\begin{document}
%
\title{Parasitic Bipolar Leakage in III-V FETs:
Impact of Substrate Architecture}
%
%
%

\author{Borna~Obradovic,
        Titash~Rakshit,
				Wei-E~Wang, 
				Dennis~Lin, 
				Niamh~Waldron,
				Nadine~Collaert, 
        and~Mark~S.~Rodder
\thanks{B. Obradovic, T. Rakshit, W.-E Wang, and M. S. Rodder are with the 
Samsung Advanced Logic Lab, Austin TX}
\thanks{D. Lin, N. Waldron and N. Collaert are with, IMEC Leuven, Belgium.}
}
\maketitle

\begin{abstract}
InGaAs-based Gate-all-Around (GAA) FETs with moderate to high In content are shown experimentally and theoretically to be unsuitable for low-leakage advanced CMOS nodes. The primary cause for this is the large leakage penalty induced by the Parasitic Bipolar Effect (PBE), which is seen to be particularly difficult to remedy in GAA architectures. Experimental evidence of PBE in In$_{70}$Ga$_{30}$As GAA
FETs is demonstrated, along with a simulation-based analysis of the PBE behavior. The impact of PBE is investigated by simulation for 
alternative device architectures, such as bulk FinFETs and FinFETs-on-insulator. PBE is found to be non-negligible in all 
standard InGaAs FET designs. Practical PBE metrics are introduced and the design of a substrate architecture for PBE suppression is elucidated. Finally, it is 
concluded that the GAA architecture is not suitable for low-leakage InGaAs FETs; a bulk FinFET is better suited for the role.
\end{abstract}


%
\IEEEpeerreviewmaketitle

\section{Introduction}
%
%
%
%
\IEEEPARstart{M}{OSFETS} of various architectures based on InGaAs have been investigated by many authors as candidates for high-performance nFETs in advanced nodes \cite{Intel},\cite{Rodwell},\cite{Kalma}. The primary attraction of the InGaAs channel is the low transport mass of the $\Gamma$ valley, and the correspondingly high electron velocity. In spite of the low charge density (also a consequence of the small
and isotropic $\Gamma$-valley mass), InGaAs devices in the quasi-ballistic limit are theoretically expected to exceed on-state
currents provided by Si-based FETs. However, the leakage characteristics of the InGaAs channel are very problematic. Due to
the small and direct bandgap, InGaAs channels are prone to Band-to-Band Tunneling (BTBT), particularly at moderate to high
voltages. As argued in \cite{New_Direction}, mobile SoC and server products require voltages in the 0.7-0.9V range. While the BTBT current itself presents a challenge (causing significant channel-to-drain GIDL leakage), the OFF-state leakage problem is made considerably worse by the Parasitic Bipolar Effect. The PBE occurs in
devices in which the channel has a poor conductive path for the extraction of holes. The holes are generated during BTBT events
in which valence electrons tunnel into the drain. The electron is swept into the drain contact, while the hole remaining in the
channel must be either conductively extracted, or eventually recombine. In GAA (or similar) FETs, there is no direct conductive
path between the channel and the substrate. Furthermore, the source-channel barrier in the off-state prevents holes from easily
diffusing from the channel into the source. Holes accumulate in the channel (being continually generated by BTBT) until
their positive charge sufficiently lowers the energy barrier. The barrier is sufficiently lowered
when the hole current over the barrier matches the BTBT hole generation current. This steady-state condition therefore has a
reduced barrier height for electrons which are diffusing from the source into the channel, increasing the off-state leakage. 
This is known as the Parasitic Bipolar effect, and has been described and observed both in Si \cite{PBE_Si} and III-V
materials \cite{PBE_InGaAs_Old}, \cite{PBE_InGaAs}. It is of
particular interest for advanced CMOS nodes because it is observed that the same device features which are needed for good electrostatic control at scaled Lg
(i.e. GAA and similar structures) also greatly exacerbate PBE, as will be shown in this paper. This makes controlling PBE of critical importance.
  
\begin{figure}[!h]
\centering
\includegraphics[width=3.0in]{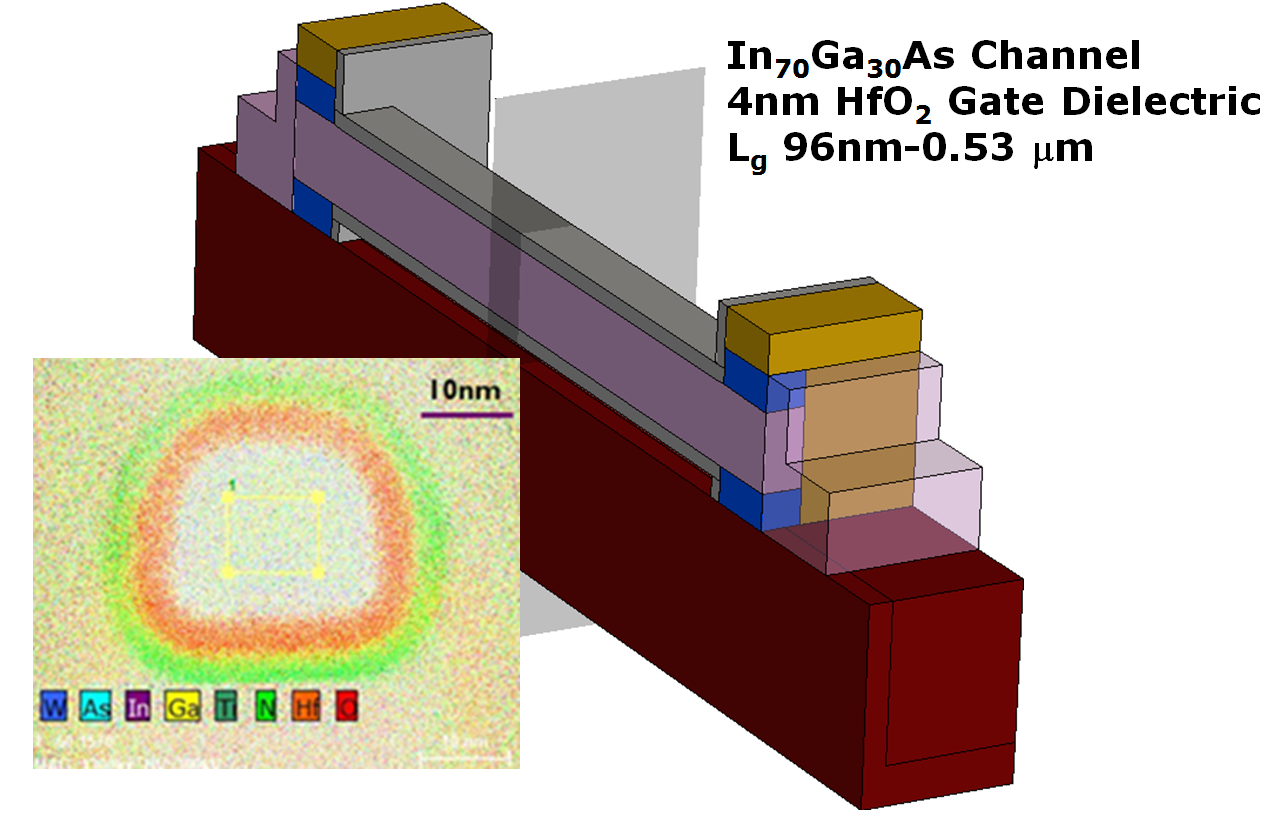}
\caption{The TEM cross-section of the GAA channels is shown. The overall (somewhat idealized) device structure
(as simulated) is illustrated for the Lg=96 nm device.}
\label{CrossSection}
\end{figure}

\section{Device Structure}

The measured device was a single-channel, rectangular nanowire GAA FET, with an In$_{70}$Ga$_{30}$As, utilizing
a novel gate dielectric (the overall process is described in detail in \cite{Kalma}). 
A typical example of the device cross-section
is illustrated in Fig. \ref{CrossSection}. Multiple channel lengths were fabricated, ranging from 'long'
channels of 0.53 $\mu$m, down to the 'short' Lg of 0.096 $\mu$m. While the latter is much longer than
the eventually desirable lengths (15-20 nm Lg for advanced CMOS nodes), it is sufficient for illustrating
the challenges posed by PBE.

\FloatBarrier

\section{Analysis of In$_{70}$Ga$_{30}$As GAA PBE}

\begin{figure}[!h]
\centering
\includegraphics[width=2.5in]{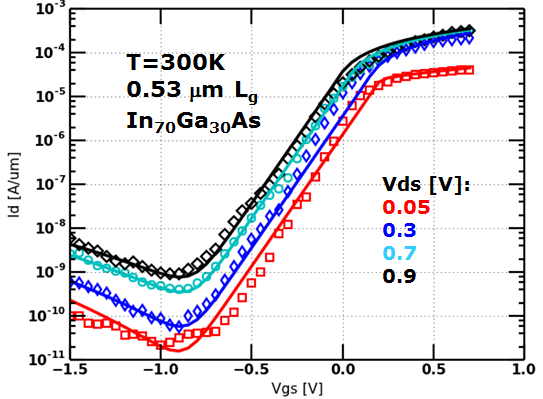}
\caption{The room-temperature Id-Vg characteristics of the long-channel In$_{70}$Ga$_{30}$As GAA are illustrated.
PBE is negligible at this length. The leakage floor is set by BTBT alone.  Symbols are measured data, lines
are simulation results.}
\label{LongChannelIMEC}
\end{figure}

The presence and magnitude of PBE is difficult to ascertain directly from single-device data. The approach
adopted in this work is to use simulation to match measured data as closely as possible, deducing
the role played by PBE from the internal properties of the simulated devices. As the first step, the long
channel device is simulated (Fig. \ref{LongChannelIMEC}). PBE is generally negligible for long channel devices
(bipolar gain scaled inversely with channel length under ideal circumstances \cite{PBE_InGaAs},
faster in non-ideal devices).
The degraded sub-threshold slope (SS) observed with the long-channel devices of Fig. \ref{LongChannelIMEC} 
(approximately 120 mV/dec) is therefore 
attributed to the presence of C$_{it}$. Simulations indicate that in the absence of  C$_{it}$, the long-channel
undoped device would have nearly ideal SS (with a fully depleted channel). This enables the calibration of C$_{it}$ for each measured device.
Additionally, the leakage tails observed in Fig. \ref{LongChannelIMEC} are attributed to pure BTBT (no PBE gain).
It should be noted that the long-channel device has the characteristic 'GIDL checkmark' shape for all V$_{DS}$ values.
The sub-threshold slope is essentially constant and mostly independent of V$_{DS}$, until the thermionic current
intercepts the BTBT current.
Comparison to simulation is used to calibrate BTBT parameters (a non-local dynamic path BTBT model (functionally 
identical to the model used in \cite{Synopsys}) is used for
the simulation; the basic parameters are from \cite{IMEC_BTBT}, with compositional and size-quantization corrections
based on Tight-Binding simulations \cite{NEMO}; additional minor adjustments to the cross-section are used to match the data (some variation in cross-section from device to device is to be expected).

\begin{figure}[!h]
\centering
\includegraphics[width=2.5in]{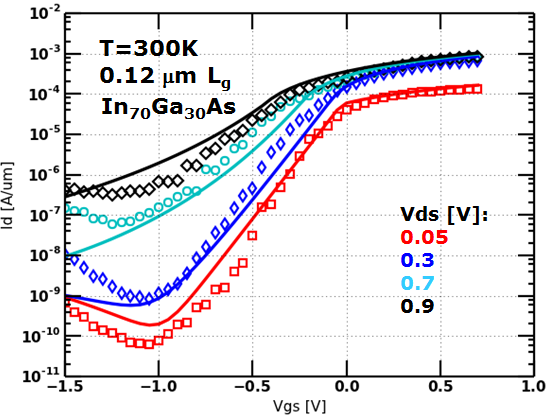}
\caption{The room-temperature Id-Vg characteristics of the short-channel In$_{70}$Ga$_{30}$As GAA are illustrated.
Symbols are measured data, lines are simulation results.}
\label{ShortChannelIMEC}
\end{figure}

The OFF-state behavior of the short-channel device (Fig. \ref{ShortChannelIMEC}) is strikingly different
than that of the long-channel device. While the 'GIDL checkmark' is still observed for low V$_{DS}$ values, there is 
a substantial increase and V$_{DS}$ dependence in the sub-threshold slope for all cases where V$_{DS}$ $>$ 0.3V. Matching this behavior in simulation
requires PBE; electrostatic degradation through D$_{it}$ does not produce the observed V$_{DS}$ behavior. Simulation also does
not indicate the presence of punch-through at this moderate channel length (which would tend to manifest as a sharp
change in SS at some point in sub-threshold; generally not a smooth increase as seen in the data). Reasonable agreement of 
simulation and data was only observed when simulating with PBE (this is actually automatic when simulating BTBT
in a substrate-decoupled channel such as GAA; suppressing PBE in simulation actually requires post-processing
BTBT current). 

\begin{figure}[!h]
\centering
\includegraphics[width=2.5in]{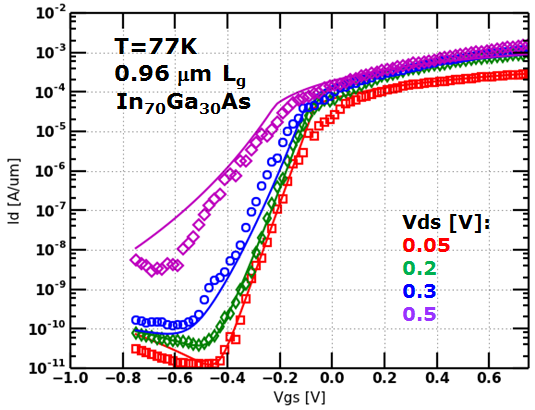}
\caption{The low-temperature (77K) Id-Vg characteristics of the short-channel In$_{70}$Ga$_{30}$As GAA are illustrated.}
\label{lowT}
\end{figure}

In order to rule out any possibility that the SS degradation with V$_{DS}$ is in fact caused by some unexpected
trap behavior (and therefore not captured in the simulation), measurements of a short device (L$_g$=96 nm) were performed at low temperature (77K). At 77K, traps near the conduction band (CB) edge are essentially unoccupied in the OFF-state,
significantly reducing the role of  D$_{it}$ on sub-threshold behavior. As can be seen in Fig. \ref{lowT}, the low-T and
room-T behavior are very similar. While the overall SS values are reduced at low-T (as is to be expected for thermionic current), the V$_{DS}$-dependent
increase is still observed. It should also be noted that the BTBT-induced leakage floor is also lower at low-T. This is a
consequence of the increased bandgap, and increased Fermi-blocking of tunneling events due to the degenerately occupied
drain. However, while the overall leakage magnitude is reduced at low-T, the basic behavior of increasing SS with V$_{DS}$
continues to be observed. As was the case at room temperature, this behavior is qualitatively reproduced by simulation 
at low-T as well. 

The match between simulation and data for the short-channel cases (Figs. \ref{ShortChannelIMEC} and \ref{lowT})
are reasonably good. In particular, the transition between the 'GIDL checkmark' and the 
more flattened Id-V$_{GS}$ behavior of the PBE-dominated high V$_{DS}$ cases is difficult to capture precisely. In the presence of PBE,
straightforward simulation indicates that all Id-Vg curves have the flattened behavior, with no uptick at very negative V$_{GS}$
values. However, recombination in the channel bulk and at the channel-dielectric interface reduces PBE gain. Interface
recombination in particular becomes strong at very negative V$_{GS}$ (when the channel holes are attracted to the gate), and eventually
suppresses PBE. When that happens, the 'GIDL checkmark' is recovered. The standard surface recombination model \cite{Synopsys} was used for
the simulations (surface recombination
velocities in the range of 1000-2000 cm/s are required for best agreement with measured data, much higher than would be
expected in a Silicon process). Even qualitative agreement of the measured and simulated data in sub-threshold could only
be accounted for with PBE. 

\begin{figure}[!h]
\centering
\includegraphics[width=2.5in]{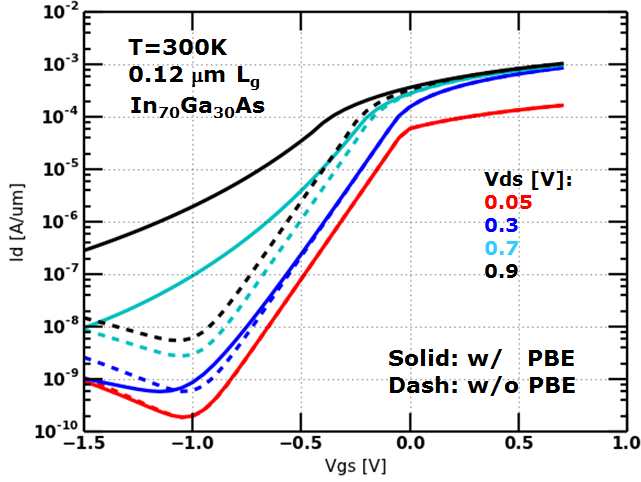}
\caption{A comparison of simulated Id-Vg curves of the short-channel In$_{70}$Ga$_{30}$As GAA are shown, with
and without PBE. The latter is only achievable in simulation, while the former is identical to the simulation
of Fig. \ref{ShortChannelIMEC}. The large increase in sub-threshold slope and leakage in general is observed to
be due to the presence of PBE. }
\label{Comparison}
\end{figure}

The impact of PBE on the device is most clearly seen in Fig. \ref{Comparison}, where simulated Id-Vg curves
for the short-channel FET are illustrated with and without PBE. The PBE-curves are identical to those of 
Fig. \ref{ShortChannelIMEC} and reasonably well-matched to data. The non-PBE curves are obtained by performing single-carrier
simulation with no BTBT, then adding a post-processed BTBT current. Given that the BTBT currents themselves are quite
low, this is tantamount to instantaneously extracting generated holes from the channel, i.e. a good approximation to 
PBE-free behavior without any structural changes (the alternative of adding a lead to the channel, or artificially 
increasing recombination, would have produced undesired simulation artifacts, and the interpretation of results would have
been consequently more difficult). It is apparent that non-PBE simulations are not even rough approximations of the
observed short-channel behavior, in spite of the good agreement at long channel. Secondly, it is also clear that
the leakage penalty of PBE is quite severe for this device; 1-2 orders of magnitude in additional leakage are
observed. 

\FloatBarrier

\section{Physics of PBE Behavior}

In the previous section it was mentioned that non-PBE BTBT leakage could be identified by the 'GIDL checkmark' shape,
whereas strong PBE tends to produce a much flatter Id-Vg curve. This qualitative difference can be used to discern 
PBE-controlled from non-PBE leakage. While the previous sections
focused on In$_{70}$Ga$_{30}$As FETs in the 96 nm-500 nm Lg range for which measured data was available, this (and subsequent) 
sections examines the behavior of devices more suited for advanced CMOS nodes. The structures considered have an Lg of 15 nm, and utilize the less
BTBT-prone In$_{53}$Ga$_{47}$As channel (an even better choice might be In$_{35}$Ga$_{65}$As, but BTBT leakage is
so dramatically reduced for this low-In channel that In$_{53}$Ga$_{47}$As serves as a better illustration of the 
challenges involved.).

\begin{figure}[!h]
\centering
\includegraphics[width=3.5in]{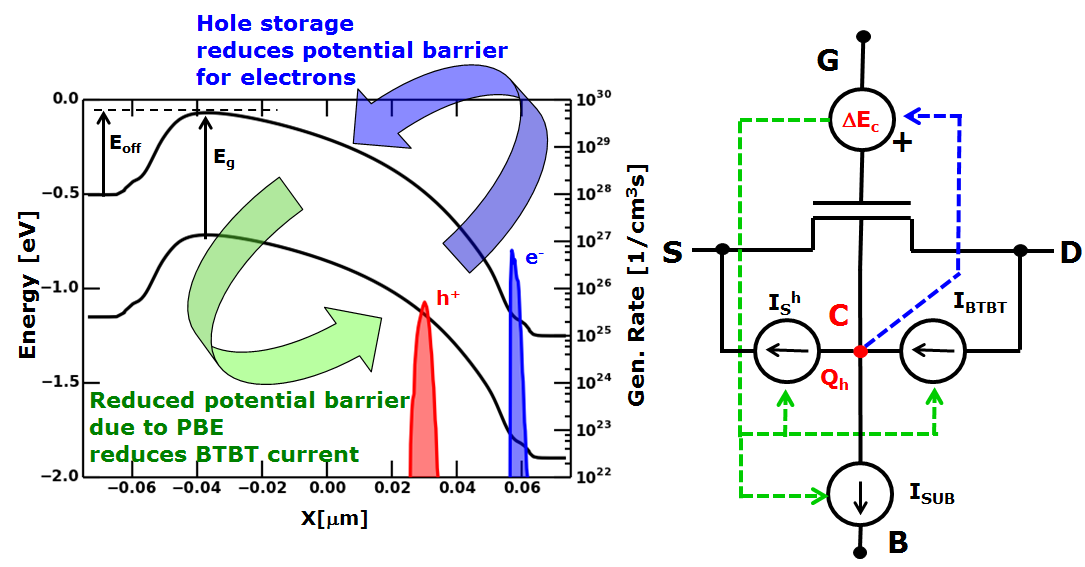}
\caption{The dual feedback mechanisms of PBE are illustrated. The first (blue) loop involves the
creation of holes in the channel via BTBT; the buildup of holes lowers the conduction band (CB) energy,
increasing electron injection. The second (green) loop couples the lowered CB to the BTBT current; lowered CB
results in decreased BTBT current. In steady state, I$_{BTBT}$, I$_S^h$, and I$_{SUB}^h$ balance, setting
the quasi-static channel hole charge Q$_h$. }
\label{PBE_feedback}
\end{figure}

\begin{figure}[!h]
\centering
\includegraphics[width=3.3in]{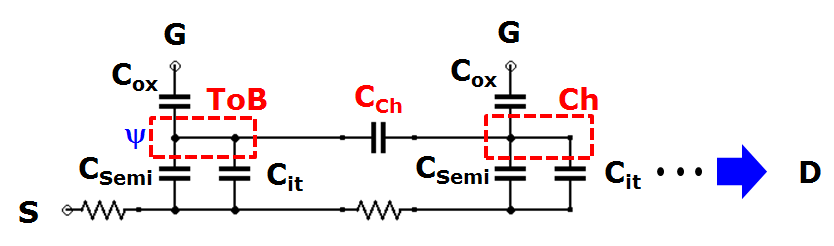}
\caption{The conceptual capacitive coupling of the ToB to the electrodes and the channel potential
is illustrated (in sub-threshold). Due to the proximity of the ToB and the accumulated channel charge,
the coupling capacitance C$_{ch}$ may be large. The capacitive coupling of the gate to the surface
potential $\psi$ is diminished
by the interface charge capacitance (C$_{it}$). With large C$_{it}$, even a relatively long channel
device (such as the measured 0.12 $\mu$m and 0.096 $\mu$m FETs) can exhibit strong PBE. This can be
thought of as an 'internal DIBL' effect, in which the role of positive drain charge is played
by the positive hole charge, but placed much closer to the ToB (as illustrated in Fig. \ref{PBE_holes}).  }
\label{circuit}
\end{figure}

The PBE mechanisms can be thought of as a dual feedback loop, as illustrated in 
Fig. \ref{PBE_feedback}. The first and well-documented loop (\cite{PBE_InGaAs_Old}, \cite{PBE_InGaAs}) 
describes the buildup of
the hole concentration in the channel, induced by the e$^{-}$/h$^{+}$ pair generation
due to BTBT at the drain side of the channel. In steady-state, the generation of holes
in the channel is balanced by the hole current across the channel-source barrier, 
the current through a substrate contact (if available), as well as bulk and surface recombination.
The equation for current continuity at the 'C' node of Fig. \ref{PBE_feedback} is then
\begin{equation}
\frac{dQ_h}{dt} = I_{btbt}(\Delta E_C) - I_s^h(\Delta E_C) -I_{b}(\Delta E_C) -\frac{\Delta Q_h}{\tau_h}
\label{Kirchhoff}
\end{equation}
\noindent where $\Delta Q_h$  is the excess hole charge and $\tau_h$ is the hole recombination time.
It is evident from Fig. \ref{PBE_feedback} and Eqn. \ref{Kirchhoff} that the various hole current
components are all functions of the CB modulation $\Delta E_C$. This modulation in turn is a function
of the accumulated hole charge, as governed by the device electrostatics.
The conceptual model for capacitive coupling of the hole charge and
the CB at the 'Top-of-the-Barrier' (ToB) is illustrated in Fig. \ref{circuit}.

\begin{figure}[!h]
\centering
\includegraphics[width=3.5in]{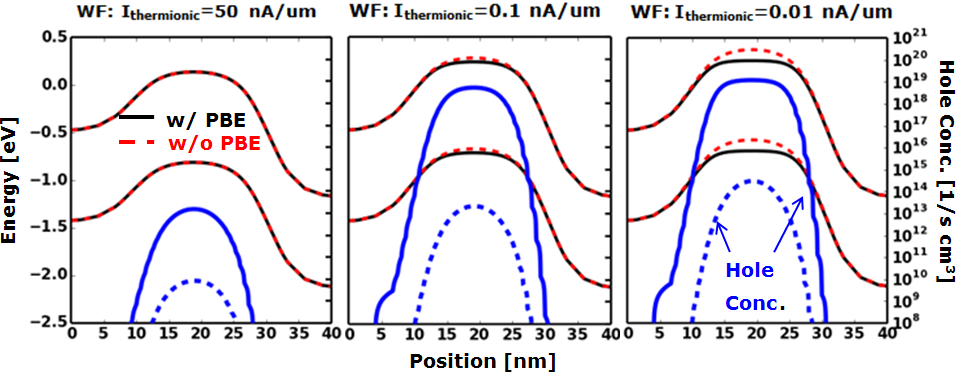}
\caption{The evolution of PBE hole storage is illustrated for varying target OFF conditions (target
thermionic I$_{OFF}$ does not include BTBT or PBE effects). As the target I$_{OFF}$ is
reduced and the source-channel barrier is increased, PBE effects become more pronounced
(due to increasing I$_{BTBT}$. This is evidenced by the increasing energy gap between the PBE
and non-PBE simulations and the increasing concentration of the stored hole charge. 
At sufficiently high Vt values, the barrier lowering stops due
to feedback to I$_{BTBT}$. }
\label{PBE_holes}
\end{figure}

The accumulation of positive charge in the channel increases the ToB potential,
but the extent to which it does so depends on the 3-D electrostatics of the device. With ideal 
electrostatic coupling of the gate to the ToB, PBE gain would be negligible (since the ToB potential
would be perfectly pinned by the gate). In real devices, however, the details of the electrostatics
must be taken into account. In this work, electrostatics are handled by 3-D Drift-Diffusion simulation.
The simulated evolution of the PBE effect as a function of the threshold voltage (expressed
as the I$_{off}$ target) is illustrated in Fig. \ref{PBE_holes}. The large positive charge
in the channel lowers the conduction band (CB) energy (also seen in Fig. \ref{PBE_holes}),
resulting in an increase in electron injection from the source. This is the 'standard'
PBE loop. However, Fig. \ref{PBE_holes} reveals a more subtle behavior: the CB lowering
begins to saturate at a certain V$_t$ (or equivalently, a certain V$_{GS}$), with only minor 
subsequent increases in the hole concentration.

\begin{figure}[!h]
\centering
\includegraphics[width=3.5in]{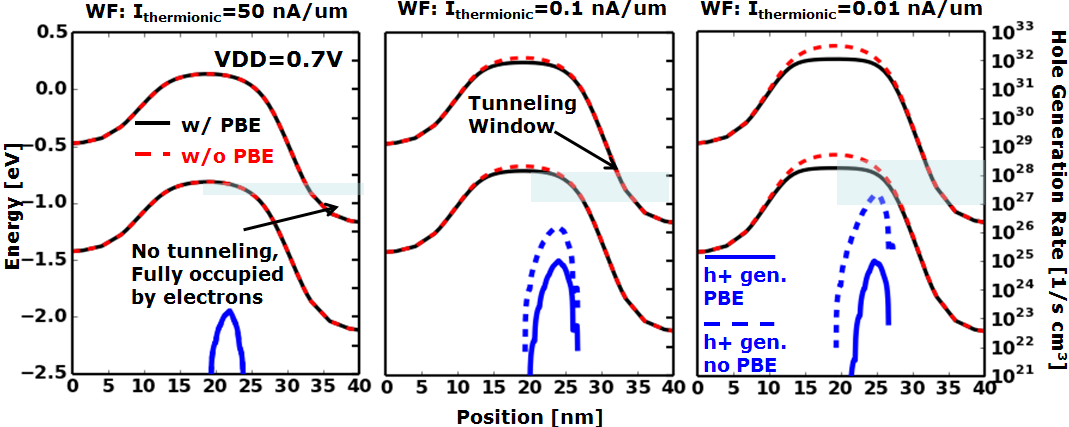}
\caption{The evolution of PBE-modulated I$_{BTBT}$ is illustrated for varying target OFF conditions (target
thermionic I$_{OFF}$ does not include BTBT or PBE effects). The I$_{BTBT}$ current is estimated by the
hole generation rate in the channel (electron generation rate in drain not shown for clarity). The hole
generation rate initially increases with decreasing target I$_{OFF}$ (standard BTBT behavior), but levels
off after a certain threshold has been reached. This is due to the feedback mechanisms coupling CB lowering
and I$_{BTBT}$.}
\label{PBE_IBTBT}
\end{figure}

This occurs in spite of the fact that I$_{BTBT}$
is expected to increase with increasing V$_t$ or increasingly negative V$_{GS}$. The reason
for this behavior can be gleaned from Fig. \ref{PBE_feedback}: the secondary feedback loop 
modulates I$_{BTBT}$ based on the shift in the CB. Specifically, as the CB is decreased
due to PBE, I$_{BTBT}$ is likewise decreased due to a reduction in the tunneling window, as
well as the effective thickness of the tunneling barrier at the drain side of the channel. This steep
reduction of I$_{BTBT}$ with increasing PBE is illustrated in Fig. \ref{PBE_IBTBT}.

As can be seen in the progression of V$_t$ values in Fig. \ref{PBE_IBTBT},
the FET with PBE exhibits a saturation of the hole generation rate with increasing
V$_t$. This is in contrast to the non-PBE device (with a substrate contact)
shown in the same Fig. \ref{PBE_IBTBT}, where the BTBT-induced hole generation rate
keeps increasing with V$_t$ (and likewise with increasing V$_{GS}$). The saturating value
of the hole concentration, CB lowering, and I$_{BTBT}$ with increasing V$_t$ or gate bias
are all consequences of the steady-state balance of the two feedback mechanisms.

\subsection{Short-Channel Prediction}
\begin{figure}[!h]
\centering
\includegraphics[width=3.0in]{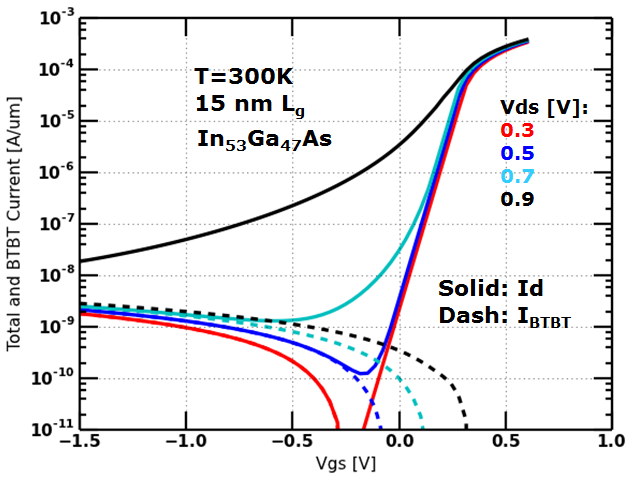}
\caption{The simulated Id-Vg behavior of a 15nm Lg In$_{53}$Ga$_{47}$As-on-insulator device is shown.
Solid lines represent the total drain current, while the dashed lines are the simulated
BTBT current. At low V$_{DS}$, the tail of the Id-Vg curve is largely set by BTBT. At higher
V$_{DS}$ (0.7V, 0.9V), however, the total channel current far exceeds the BTBT current.  }
\label{Id_Lg15nm}
\end{figure}

The predicted current behavior of the 15nm FET is discussed next.
The Id-Vg characteristics are illustrated in Fig. \ref{Id_Lg15nm}.

\begin{figure}[!h]
\centering
\includegraphics[width=3.0in]{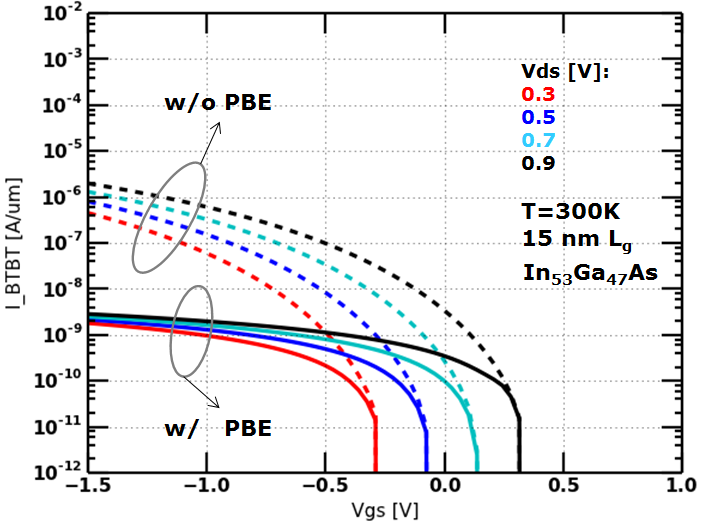}
\caption{The simulated Id-Vg and I$_{BTBT}$behavior of a 15nm Lg InGaAs-on-insulator device is 
illustrated. Solid lines represent the BTBT current in the presence of PBE, while the dashed lines
are what the BTBT current would have been if holes could be instantaneously extracted from the channel
(completely suppressing PBE). A significant reduction in the actual BTBT current is observed due
to the second PBE feedback mechanism.  }
\label{BTBT_wPBE}
\end{figure}

The large discrepancy seen between the total drain current and the BTBT current
for moderate to large drain biases observed in Fig. \ref{Id_Lg15nm} is of course due to
PBE (as also reported in \cite{PBE_InGaAs_Old} and \cite{PBE_InGaAs}). The standard metric for bipolar gain is defined as
\begin{equation}
\beta = \frac{\Delta I_D}{I_{BTBT}}
\end{equation}
\noindent where $\Delta I_D$ is the increase in drain current due to PBE, while I$_{BTBT}$ is
the BTBT current itself. The bipolar amplification factor $\beta$ is often interpreted as the `leakage penalty`
incurred due to PBE. However, this is not entirely correct since the presence of PBE alters
BTBT itself (the second feedback loop). This is illustrated in the case of the 15nm device 
in Fig. \ref{BTBT_wPBE}.

Due to the significant discrepancy between the BTBT currents with and without PBE, a more
direct metric for the `PBE leakage penalty` is defined here as follows:
\begin{equation}
\Gamma=\frac{I_D^{PBE}}{I_D^{no\ PBE}}
\label{Gamma}
\end{equation}

\begin{figure}[!h]
\centering
\includegraphics[width=3.0in]{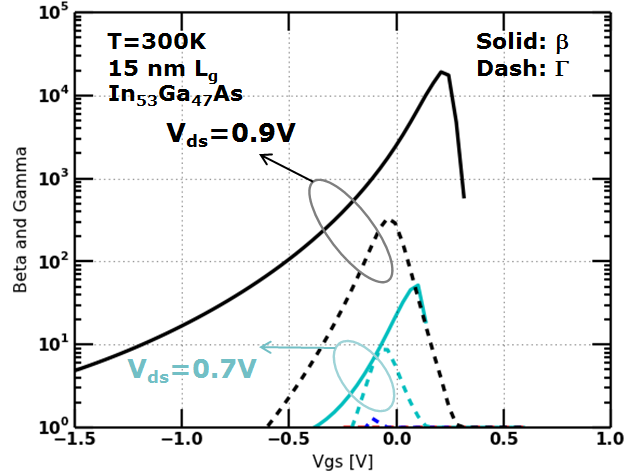}
\caption{The bipolar gains $\beta$ and $\gamma$ are illustrated for the 15nm Lg InGaAs-on-Insulator
device. Large gains are achieved only for moderate and high V$_{DS}$ values (0.7V and 0.9V), and are
non-monotonic with V$_{GS}$. The true bipolar gain penalty $\gamma$ is generally somewhat lower
than the standard $\beta$ metric would indicate. }
\label{Beta_and_Gamma}
\end{figure}

\noindent The $\Gamma$ gain of Eqn. \ref{Gamma} is the ratio of the drain current with and without PBE.
The BTBT currents are of course very different for the two cases, as are the drain currents. The case
with PBE corresponds to a real (measured or simulated) device, whereas the no-PBE case is an imaginary
device in which the holes accumulating in the channel are instantly extracted. Given this definition,
$\Gamma$ is necessarily a simulation-based metric. The behavior of $\beta$ and $\Gamma$ as functions
of applied bias are illustrated in Fig. \ref{Beta_and_Gamma}.

As can be seen in Fig. \ref{Beta_and_Gamma}, bipolar gain $\beta$ is quite large, easily 
reaching into the thousands at high V$_{DS}$. At low V$_{DS}$, $\beta$ is negligible as expected, simply due
to the fact that BTBT-driven hole generation is too low for any significant accumulation in the channel.
The V$_{GS}$ behavior of $\beta$ is non-monotonic. At high V$_{GS}$ values, the source-channel barrier is
reduced or suppressed entirely and holes can escape from the channel into the source. Thus, the steady-state hole
concentration is low. At very negative V$_{GS}$ the feedback between I$_{BTBT}$ and the barrier height
is pinning I$_D$ to a nearly constant value, suppressing further bipolar gain. In the intermediate
region near the OFF-state of the transistor, bipolar gain is near maximum. The leakage penalty 
gain $\gamma$ likewise peaks near the target OFF-state, albeit at lower values than $\beta$.
At high V$_{DS}$, the true leakage penalty induced by PBE is seen to be approximately 100. Of course,
this value is very device-structure specific. It can also be seen from Fig. \ref{Beta_and_Gamma} that
$\Gamma$ is actually less than unity for strongly negative V$_{GS}$, which is another way of saying that
the leakage current in the presence of PBE is lower than it would have been without PBE. This is also 
a consequence of the pinned value of BTBT current in the presence of PBE (pinned when operating at 
strongly negative V$_{GS}$). The relatively flat leakage tail in the presence of PBE is necessarily lower
than the 'GIDL checkmark' at sufficiently negative V$_{GS}$. However, this should not be interpreted to mean
that PBE is preferable to PBE-free operation; leakage in the standard operating regions of V$_{GS}$ is significantly
worse in the presence of PBE (as shown in Fig. \ref{Beta_and_Gamma}).

\begin{figure}[!h]
\centering
\includegraphics[width=2.5in]{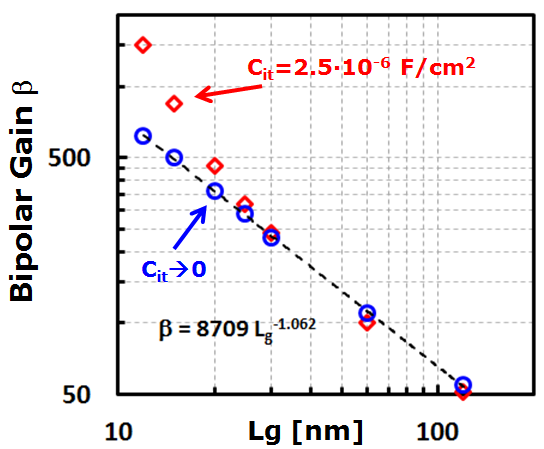}
\caption{The dependence of bipolar gain $\beta$ on gate length and electrostatics for the FinFET-on-insulator
device is illustrated. Along
with varying gate length, two flavors of the FinFET are shown: with zero C$_{it}$ (zero D$_{it}$) and with a large value
of C$_{it}$ (large D$_{it}$). The devices are otherwise identical. The zero-C$_{it}$ device exhibits roughly inverse-linear
scaling of $\beta$ with L$_g$ across all lengths considered, as would be expected for ideal bipolar gain.
The device with large C$_{it}$ exhibits inverse-linear scaling of gain only in the long-L$_g$ regime; below ~30nm
L$_g$, bipolar gain is seen to increase rapidly and does not follow the expected inverse-linear behavior.
This is a consequence of the loss of electrostatic control of the ToB region, as illustrated in Fig. \ref{circuit}.}
\label{Beta-Lg}
\end{figure}

It was previously mentioned that gate electrostatics play a key role in determining the bipolar gain. Based on
the capacitive voltage division suggested by Fig. \ref{circuit}, it is clear that control of the surface
potential at the ToB is easily
asserted by the 'deep' channel potential via C$_{ch}$ if C$_{it}$ is large compared to C$_{ox}$. This coupling
is a DIBL-like effect. Instead of remote positive charge in the drain influencing the ToB potential (which of 
course still takes place), there is a barrier lowering effect of the much more closely placed hole charge.
Due to the proximity of the hole charge, the effective coupling capacitance of the holes to the ToB is larger
than would be used to model the DIBL effect. We should therefore expect an electrostatics-induced departure from
the usual $1/L_{g}$ scaling of bipolar gain when channel lengths approach the electrostatic limit imposed by the
capacitive coupling to hole charge. 
This is the case of the measured In$_{70}$Ga$_{30}As$ described in the first section of the paper, in which
even a moderately long-channel device (120 nm) is experiencing significant PBE. In Fig. \ref{Beta-Lg},
this effect is illustrated by examining the L$_g$-dependence of $\beta$ for two cases: with and without C$_{it}$.
The case of no C$_{it}$ capacitor (i.e. C$_{it} \rightarrow 0$)  has the best electrostatics possible within the geometry of the FinFET-on-insulator. Bipolar gain
is seen to scale very nearly with the expected inverse-L$_g$ law (as reported in \cite{PBE_InGaAs_Old} and \cite{PBE_InGaAs}). However, the finite
C$_{it}$ case shows strong deviation from the inverse-L$_g$ law below ~30 nm L$_g$. As suggested by Fig. \ref{circuit}, the
presence of a large C$_{it}$ reduces the coupling of the surface potential at the ToB to the gate, diminishing the SCE performance in general, but more specifically
allowing the ToB potential to be more closely controlled by the deep channel potential. This makes the source-channel barrier
more susceptible to potential modulation by the injected hole charge, resulting in increased PBE gain. It should therefore
be expected that FETs operating near the edge of Vt-rolloff (i.e. advanced CMOS devices pushed to the limit of L$_g$ scaling).
will have higher bipolar gains than would have been predicted based on inverse-L$_g$ scaling alone.

\section{Impact of Substrate Architecture}

In the previous sections it has been argued that the rate at which BTBT-generated holes can be extracted
from the InGaAs channel is a key determinant of the magnitude of PBE gain, and thereby the OFF-state
leakage. In this section, the choice of substrate architecture is shown to be critical importance in
controlling PBE. The standard architectures usually considered for advanced nodes fall into several 
categories:
\begin{enumerate}
\item{FinFET on semiconductor substrate ('bulk' substrate)}
\item{FinFET on insulating substrate}
\item{GAA configuration, such as a NW or Nanosheet ('wide NW')}
\end{enumerate}

\begin{figure}[!h]
\centering
\includegraphics[width=2.75in]{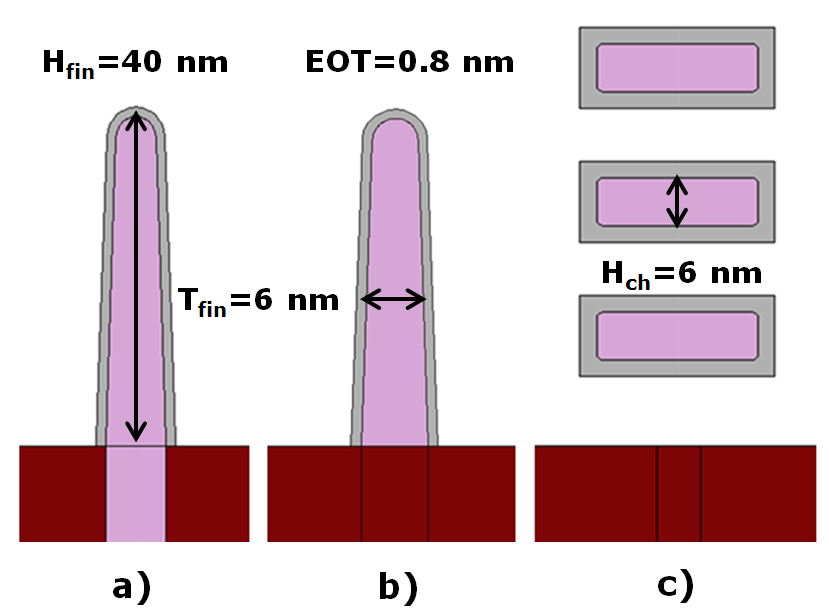}
\caption{The device architectures considered in this study are illustrated.
Fin-on-Bulk (device 'a') has an InGaAs channel and a semiconductor substrate. The semiconductor
may be InGaAs, or it may be InP or InAlAs (or other). A remote well contact connected to the
substrate material is assumed. Device 'b' is the Fin-on-Insulator case. No conductive substrate contact is available.
Device 'c' is a stack of Nanosheets (also referred to as 'wide nanowires'), a GAA architecture with
no substrate contact.
 }
\label{Devices}
\end{figure}

\begin{figure}[!h]
\centering
\includegraphics[width=3.0in]{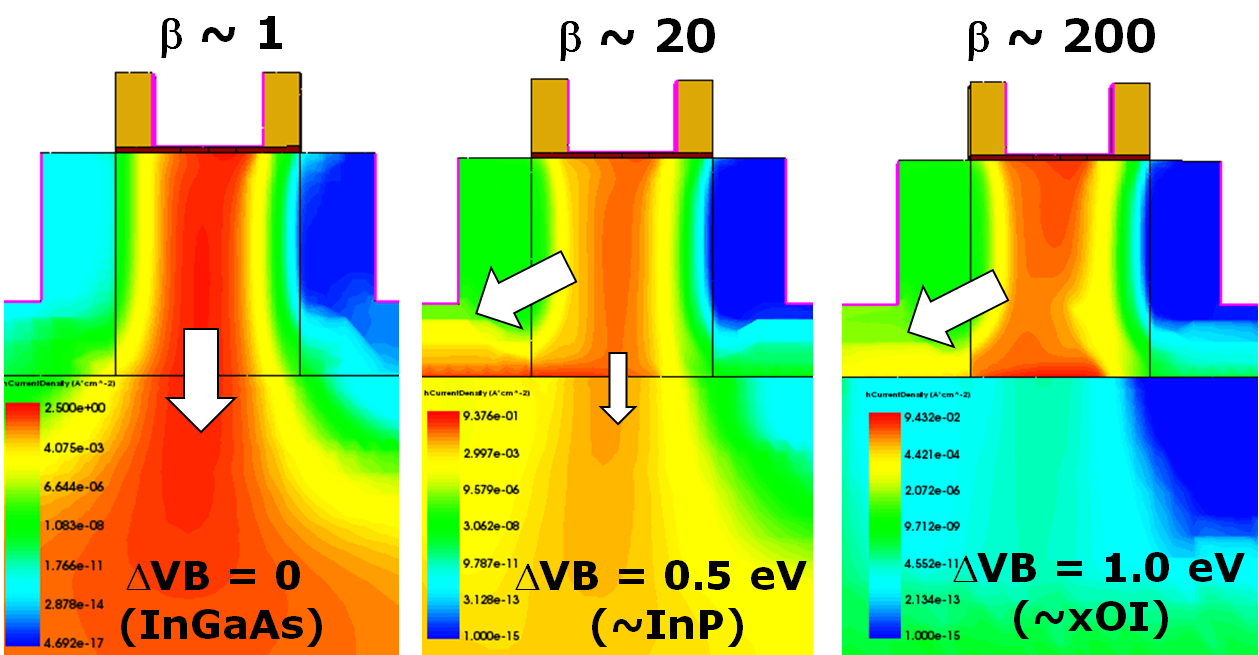}
\caption{The hole current paths in the OFF-state are illustrated for three FinFET substrate types.
In the absence of the VB offset at the channel-substrate interface (no barrier for holes) hole current
is seen to be entirely through the substrate. With increasing offset, hole current must flow across the
channel-source barrier, necessitating the increase of the channel potential. Bipolar gain is seen to increase
with increasing VB offset to the substrate. }
\label{Barrier}
\end{figure}

\begin{figure}[!h]
\centering
\includegraphics[width=3.0in]{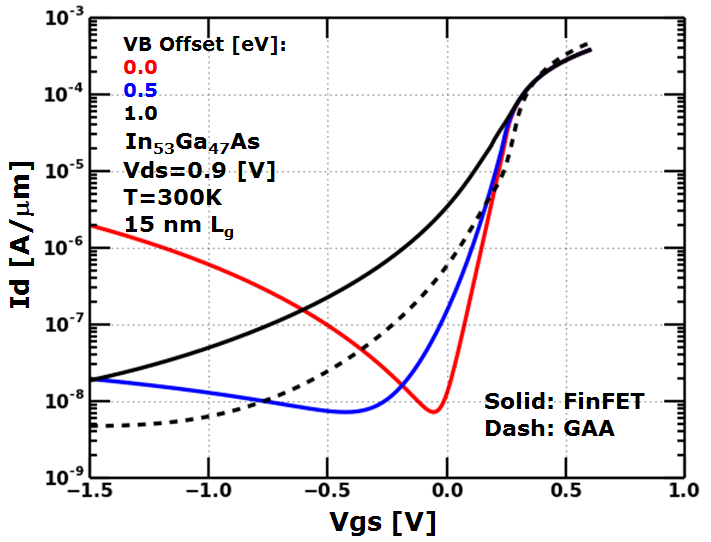}
\caption{The high-Vd Id-Vg characteristics of FinFETs and GAA structures are shown. The FinFET has 
three levels of valence barrier offset relative to the substrate: 0, 0.5eV, and 1 eV. The GAA device
has no connection of the channel to the substrate. Increasing the barrier height has the basic effect
of increasing PBE. The zero-barrier FinFET has BTBT-limited leakage current. 
The GAA device exhibits somewhat lower leakage currents than the high-barrier FinFET
due to the improved electrostatics (at equal Lg).}
\label{Barrier_IV}
\end{figure}

Of the three choices listed, only the `FinFET on semiconductor` choice provides a conductive path
to the substrate. However, even in the case of the bulk substrate, the hole extraction path may be partially
blocked by an energy barrier. The barrier is a consequence of the valence band (VB) mismatch (denoted as $\Delta$VB)
between the channel
material (in this case In$_{53}$Ga$_{47}$As) and the substrate material. Common choices for the substrate include
InP and In$_{52}$Al$_{48}$As, both of which are lattice-matched to In$_{53}$Ga$_{47}$As. The former has a VB offset
of approximately 0.5 eV relative to In$_{53}$Ga$_{47}$As, while the latter has an offset of 0.3 eV. Both are 
multiple kT at room temperature, and unless the channel/substrate interface is degenerately doped to enable channel-substrate
tunneling, both barriers are expected to contribute to PBE. This is illustrated by simulation in Fig. \ref{Barrier}.
As seen in Fig. \ref{Barrier}, channels with matched substrates (VB=0) have a direct path for minority
holes to move into the substrate. No increase in channel potential is required, and the bipolar gain (shown at V$_{DS}$=0.7V)
is negligible. As the barrier height is increased to 0.5 eV (InP) or higher (approximating an insulating substrate),
the hole current path through the substrate is suppressed. Bipolar gains even at low V$_{DS}$ are in the hundreds.
The complete Id-Vg characteristics of devices with various substrates are shown in Fig. \ref{Barrier_IV}. 

\begin{figure}[!h]
\centering
\includegraphics[width=3.0in]{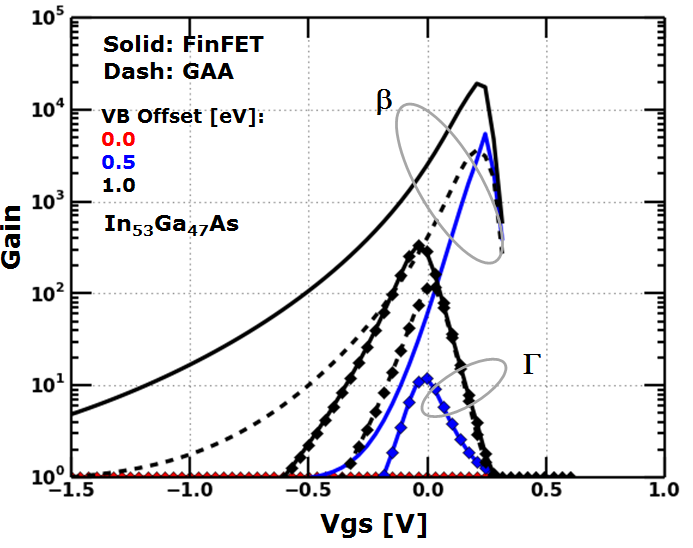}
\caption{Bipolar gain metrics $\beta$ and $\Gamma$ are illustrated for varying-barrier FinFETs and the
GAA device. While the gains are negligible for the zero-barrier device, both $\beta$ and $\Gamma$ quickly
increase with barrier height. A barrier of 1eV is essentially infinite (negligible current flows through
the substrate). Even the intermediate barrier \(0.5 eV\) shows significant bipolar gain.}
\label{Barrier_Gain}
\end{figure}

As illustrated in Fig. \ref{Barrier_IV}, increased barrier height results in the typical PBE-induced
increase of sub-threshold slope, and generally higher leakage across most of the OFF-regime. Even with an InP
substrate ($\Delta$VB=0.5 eV) the degradation in sub-threshold characteristics is noticeable. Deep in the 
accumulation region, leakage with strong PBE is actually lower than that of the no-PBE case (due to feedback-induced
suppression of BTBT). However, in the normal operating regime for CMOS circuits, the high-barrier (and therefore high PBE)
cases exhibit significantly higher leakage than the matched substrate case. This is further highlighted by the
bipolar gains metrics shown in Fig. \ref{Barrier_Gain}.

The bipolar gains $\beta$ and $\Gamma$ shown in Fig. \ref{Barrier_Gain} both show the strong effect
of the substrate barrier. Bipolar gains with the InP substrate are in the hundreds (near the nominal OFF condition, 
with V$_{DS}$=0.9V),
while the true leakage penalty is about 10. The FinFET-on-insulator device has a leakage penalty of several hundred.
In both Fig. \ref{Barrier_IV} and \ref{Barrier_Gain}, it is apparent that the GAA Nanosheet device has lower gains
(both $\beta$ and $\Gamma$) than the corresponding FinFET-on-insulator, in spite of having no extraction path
for holes (just like the FinFET-on-insulator case). This is a consequence of the somewhat improved electrostatics of the
GAA device, relative to the bulk FinFET. The source-channel barrier of the GAA case is less perturbed by an increased
channel potential (increased due to the presence of BTBT-induced holes), resulting in a smaller PBE gain. Referring
to Fig. \ref{circuit}, the ratio of the series combination of C$_{ox}$ and C$_{it}$ to C$_{ch}$ is better in the GAA FET than
in the Fin-on-insulator FET.

It should be noted that the electrostatic benefit of a 20 nm wide GAA nanosheet over a FinFET is quite modest
(much better improvement is available with a full NW structure, but this is not considered here), and the immunity
to PBE is only slightly improved. It can therefore be concluded that a GAA device in the 15nm L$_g$ range
has a leakage penalty on the order of 100X relative to a true bulk FinFET (at the overdrive V$_{DS}$ voltage of 0.9V). This 
is certainly the case with the In$_{70}$Ga$_{30}$As GAA NW FETs of Fig. \ref{ShortChannelIMEC}. Even though the device
appears to be quite long on the electrostatic length scale, a large C$_{it}$ due to interface traps (high D$_{it}$) reduces the
coupling of the gate to the top of the barrier. In addition to standard SCE degradation, this degraded coupling is
also making the device much more susceptible to PBE.

\begin{figure}[!h]
\centering
\includegraphics[width=3.0in]{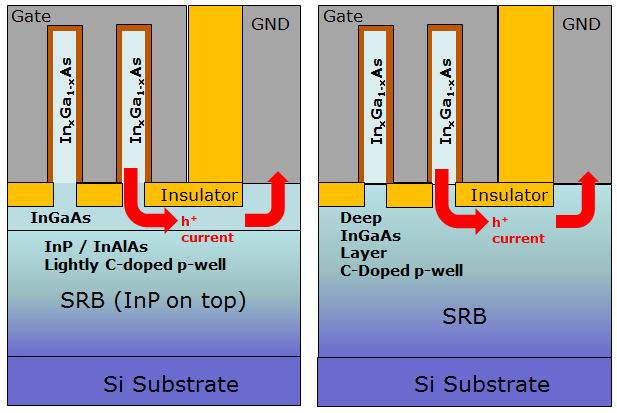}
\caption{Two options for a a contacted, zero-offset substrate are illustrated. Both options
provide near-zero bipolar gain. They differ in how they attempt to deal with sub-fin leakage, commonly
associated with Fin-on-Bulk architectures. The structure of the left figure utilizes a thin InGaAs
layer for hole extraction, with a wide-bandgap semiconductor underneath. The latter provides a large
CB offset, suppressing the flow of electrons in the deep bulk. The structure in the right figure 
has only InGaAs in the substrate (other than the obligatory SRB).}
\label{Bulk_Substrates}
\end{figure}

As shown in Figs. \ref{Barrier_Gain}, all architectures other than FinFETs on zero-VB-offset substrates
exhibit a leakage penalty, particularly under overdrive conditions. The only obvious solution to the PBE problem 
(short of using a large-bandgap channel material to suppress BTBT) is the FinFET with a substrate architecture shown in Fig. 
\ref{Bulk_Substrates}. Two flavors are illustrated, but both have two common features: the material directly 
under the fins (well top) exactly matches the channel, and a (remote) contact is placed in the well top. This ensures
that a conductive path for hole extraction exists, with no barrier that would result in a charge pileup in potential
shifts. The well top is the final layer of a thick SRB, which may include a wide-bandgap material as a subfin-leakage
suppressor (or not, in which case doping is used for subfin leakage suppression). This is in fact very similar to the
standard well contact scheme used for Si. It is a consequence of the high OFF-state leakage of narrow-bandgap materials
that this may in fact be the only architecture possible for a low-power mobile SOC. For all its benefits, full GAA with 
a narrow bandgap material 
has a seemingly incurable PBE problem.

\FloatBarrier
\section{Summary and Conclusion}

This paper presents evidence of strong PBE gains in GAA devices that utilize narrow-bangdap InGaAs. Measured data
of moderate-length GAA devices exhibit a high level of leakage not consistent with BTBT alone. Furthermore, the 
V$_{DS}$ and V$_{GS}$ dependence of the leakage tails is not consistent with either pure BTBT leakage or punchthrough, but is 
well explained by PBE in conjunction with BTBT (through simulation), both at room and low temperature. Some aspects of the physics of PBE
are elucidated upon. In particular, the dual-feedback mechanism which is responsible for the flattening of the
Id-Vg curves in the OFF-state of devices with strong PBE is explained. It is also emphasized that this flattening is much more
detrimental to FET behavior than the usual BTBT leakage. While the latter sets a leakage floor, PBE in conjunction
with BTBT increases the sub-threshold slope in the near-OFF regime (the tail is not exponential, but has an increasing
`local` SS as the device is pushed further toward the OFF state). It is argued that this behavior is fundamental
to GAA FETs, due to the absence of a conductive path between the channel body and the substrate. Low-bandgap materials
such as In$_{53}$Ga$_{47}$As (simulated in paper) or In$_{70}$Ga$_{30}$As (data and simulation in paper) are therefore not
suitable for low-power SoC applications if GAA architectures are to be employed. Similar conclusions apply to FinFET-on-insulator
architectures, since conductive coupling between the channel and the substrate does not exists. Even the case of the
FinFET on semiconductor substrate is potentially problematic. The standard approach of using a wide-bandgap semiconductor 
(such as InP or InAlAs) as the substrate for InGaAs FinFETs is also found to lead to PBE gain, albeit more moderate than the
GAA or the xOI case. The cause of the PB effect in this case is the valence band offset between the InGaAs channel and the
InP (or InAlAs) substrate. This offset presents a barrier for holes, resulting in hole accumulation in the channel (i.e. the
standard PBE mechanism). The only substrate option which eliminates PBE is found to be one in which the channel and the substrate
material are matched (or at least their valence band edges are aligned) and a well contact is placed into the matched well material (possibly in a remote cell which shares the
substrate material). Thus, it would appear that a bulk FinFET architecture is preferable to GAA for low-leakage InGaAs devices.


%



\ifCLASSOPTIONcaptionsoff
  \newpage
\fi

\end{document}